\documentclass[twocolumn]{aastex63}

\received{June 1, 2019}
\revised{January 10, 2019}
\accepted{\today}
\submitjournal{ApJ}

\shorttitle{Tidal Interaction between the UX Tauri Disk A/C System Revealed by ALMA}
\shortauthors{Zapata et al.}
\usepackage{graphicx}

\begin{document}

\title{Tidal Interaction between the UX Tauri A/C Disk System Revealed by ALMA}

\correspondingauthor{Luis A. Zapata}
\email{l.zapata@irya.unam.mx}

\author{Luis A. Zapata}
\affil{Instituto de Radioastronom\'\i a y Astrof\'\i sica, Universidad Nacional Aut\'onoma de M\'exico, P.O. Box 3-72, 58090, Morelia, Michoac\'an, M\'exico}

\author{Luis F. Rodr\'\i guez}
\affil{Instituto de Radioastronom\'\i a y Astrof\'\i sica, Universidad Nacional Aut\'onoma de M\'exico, P.O. Box 3-72, 58090, Morelia, Michoac\'an, M\'exico}
\affil{Mesoamerican Centre for Theoretical Physics, Universidad Aut\'onoma de Chiapas, Carretera Emiliano Zapata Km. 4 Real del Bosque, 29050 Tuxtla Guti\'errez, Chiapas, M\'exico}

\author{Manuel Fern\'andez-L\'opez}
\affiliation{Instituto Argentino de Radioastronom\'\i a (CCT-La Plata, CONICET; CICPBA), C.C. No. 5, 1894, Villa Elisa, Buenos Aires, Argentina}

\author{ Aina Palau}
\affil{Instituto de Radioastronom\'\i a y Astrof\'\i sica, Universidad Nacional Aut\'onoma de M\'exico, P.O. Box 3-72, 58090, Morelia, Michoac\'an, M\'exico}

\author{Robert Estalella}
\affil{Dep. de F\'{\i}sica Qu\`antica i Astrof\'{\i}sica, Institut de Ciencies del Cosmos, Universitat de Barcelona, IEEC-UB, Marti i Franques 1, 08028 Barcelona, Spain}

\author{Mayra Osorio}
\author{Guillem Anglada}
\affil{Instituto de Astrof\'\i sica de Andaluc\'\i a (CSIC), Glorieta de la Astronom\'\i a s/n E-18008 Granada, Spain}

\author{Nuria Huelamo}
\affil{Centro de Astrobiolog\'\i a (CSIC-INTA), Camino bajo del Castillo s/n, 28692 Villanueva de la Ca\~nada, Madrid, Spain}




\begin{abstract}

We present sensitive and high angular resolution ($\sim$0.2-0.3$''$)  (sub)millimeter (230 and 345 GHz) continuum 
and CO(2$-$1)/CO(3$-$2) line archive observations of the disk star system in UX Tauri carried out with ALMA (The Atacama Large Millimeter/Submillimeter Array). 
These observations reveal the gas and dusty disk surrounding the young star UX Tauri A with a large signal-to-noise ratio ($>$400 in the continuum and $>$50 in the line), 
and for the first time is detected the molecular gas emission associated with the disk of UX Tauri C (with a size for the disk of $<$56 au). No (sub)millimeter continuum
emission is detected at 5$\sigma$-level (0.2 mJy at 0.85 mm) associated with UX Tauri C.
For the component UX Tauri C, we estimate a dust disk mass of $\leq$ 0.05 M$_\oplus$.
Additionally, we report a strong tidal disk interaction between both disks UX Tauri A/C, separated 360 au in projected distance.
The CO line observations reveal marked spiral arms in the disk of UX Tauri A  and an extended redshifted stream of gas 
associated with the UX Tauri C disk.  No spiral arms are observed in the dust continuum emission of UX Tauri A. 
Assuming a Keplerian rotation we estimate the enclosed masses (disk$+$star) from their radial velocities in 
1.4 $\pm$ 0.6 M$_\odot$ for UX Tauri A, and  70 $\pm$ 30 / $\sin i$ Jupiter masses for UX Tauri C 
 (the latter coincides with the mass upper limit value for a brown dwarf).
 The observational evidence presented here lead us to propose that UX Tauri C is having 
 a close approach of a possible wide, evolving and eccentric orbit around the disk of UX Tauri A causing the formation of spiral arms and
 the stream of molecular gas falling towards UX Tauri C. 


\end{abstract}



\section{Introduction} \label{sec:intro}

One of the natural consequences of the angular momentum conservation and energy dissipation 
in the formation of the stars are circumstellar disks \citep{bat2018}. To date, dozens of circumstellar disks 
have been imaged employing different observational techniques.  In the Orion Nebula Cloud, for example,  
many of these disks have been revealed through resolving the scattered light and silhouette images of disks
using observations of the {\it Hubble Space Telescope (HST)} with an 
angular resolution of 0.1$''$ \citep{ball2000,ode2008}.  Another key technique that is used for revealing disks is the direct 
imaging of the centimeter, millimeter, and submillimeter thermal continuum and line emission utilizing interferometric 
aperture synthesis technique \citep{ans2017, will2011, rui2018, zap2017, hua2018, kur2018A, van2019, oso2016}.  Some of these sensitive 
submm observations have revealed possible encounters between close-by circumstellar disks \citep{van2019, kur2018A, aki2019, dai2015}. 
The CO(2$-$1) line and 1.25 mm continuum emission revealed clear signatures of tidal interactions 
with spiral arms and extended stream-like emission in the disk system AS 205 \citep{kur2018A}.
Given that most of the stars are formed in binary or multiple systems \citep{duc2013}, more cases of 
disks interaction might be revealed with more ALMA sensitive observations.

There are a large number of works \citep[e.g.][]{ada1989,lau1996} that explain the spiral arms in gaseous and collisionless self-gravitating disks
 under the classical non-axisymmetric density perturbations proposed for the first time to galaxies \citep{liu1964}. More recently, some other works
 using hydrodynamical simulations had generated spiral arms by the presence of a massive companion \citep{kle2012,zhu2015}. 
 For example, \citet{zhu2015} found that depending on the mass of the planet, the arms in the circumstellar disks could become more open 
 with a more massive perturber. There are also some works that propose the spiral arms as gravitational instabilities, see \citet{lod2004}. 
 However, in these works using gravitational instabilities, there are an apparent large number of spiral arms than those observed in disks. 

UX Tauri is a quadruple T Tauri star system located in the Taurus molecular cloud 
at a distance of 139 $\pm$ 2 pc \citep{bai2018}\footnote{However, we found a difference of almost 6 pc
between UX Tauri A and C from the GAIA parallaxes (7.15$\pm$01 and 6.85$\pm$0.1 mas). As we will
see these components should be very close to each other because there is a small difference in the radial velocities 
between their molecular disks ($\leq$ 1 km s$^{-1}$) and we were able to resolve the molecular gas
that is associated with an interaction. As these components (A and B) are in a binary system,  
Gaia DR2 possibly did not account for the binary motion when calculating its parallax. This is a similar case
as in AS 205 system \citep{kur2018A}. Thus, we consider the distance of UX Tauri A and C being 
the same for both.}. 
The system consists of UX Tauri A,
the main star, UX Tauri B separated by 5.6$''$ to its west, and UX Tauri C, a southern 
companion only separated by 2.6$''$ or 360 au \citep{csp2017}. UX Tauri B is itself a close binary 
system separated by 0.1$''$ \citep{cor2006}. 
The components of the system, 
UX Tauri A and C, are classified with a spectral type of K2V and M5, and stellar masses
of 1.3 $\pm$ 0.04 M$_\odot$ and 160 $\pm$ 40 Jupiter masses, respectively \citep{kra2009}. 
Analysis of the SED of UX Tauri A suggests
the presence of a pre-transitional disk with significant near-infrared excesses, which indicates 
the presence of an optically thick disk with an inner gap (R$_{wall}$) of ~56 au \citep{esp2007}.  
Subsequent sensitive Submillimeter Array (SMA) continuum observations at 345 GHz resolved angularly 
the dusty disk surrounding UX Tauri A with an angular resolution of 0.3$''$ \citep{and2011}. 
These submillimeter observations imaged the circumstellar disk inner gap around UX Tauri A and
estimated a size of 25 au from the central star (smaller than that estimated by the SED).  
Near-IR  images obtained with the Subaru Telescope revealed a 
strongly polarized circumstellar disk surrounding UX Tauri A, 
which extends to 120 au and inclined 46$^\circ$ $\pm$ 2$^\circ$ \citep{tan2012}.

We note that SMA and VLA (Very Large Array) continuum observations could not
detect its companion (UX Tauri C) at a 5$\sigma$ level, 7.5 mJy (0.85 mm) for the SMA \citep{and2011} and 20 $\mu$Jy (3.6 cm) for the VLA \citep{zap2017}.
This companion is only reported in the near- or mid- infrared \citep{cor2006,mac2006}.  At this point, the gas and dust disk associated with UX Tauri C has also not been detected 
 at (sub)millimeter wavelengths \citep{and2011}. One reason that the disk is not detected at (sub)millimeter wavelengths maybe is because
 its circumstellar dusty disk is very faint and small. Neither the dusty nor the gas disk of UX Tauri B have been detected at submillimeter wavelengths.  
  
 In this study we present new sensitive and high angular resolution ($\sim$0.2--0.3$''$) archive ALMA (Atacama Large
 Millimeter/Submillimeter Array) CO(2$-$1)/CO(3$-$2) observations that reveal a strong 
 interaction between the disks of UX Tauri A and C, separated only by about 360 au. 
 This interaction may cause the marked spiral arms in the disk UX Tauri A  and an extended stream of gas emission in the UX Tauri C disk.
 These observations additionally reveal for the first time the gas disk and its kinematics related with UX Tauri C.
 Since neither the dusty and gas of UX Tauri B have been detected in these submillimeter ALMA observations, we will focus on the A and C components. 
 
 \begin{figure*}[ht!]
\epsscale{1.165}
\plottwo{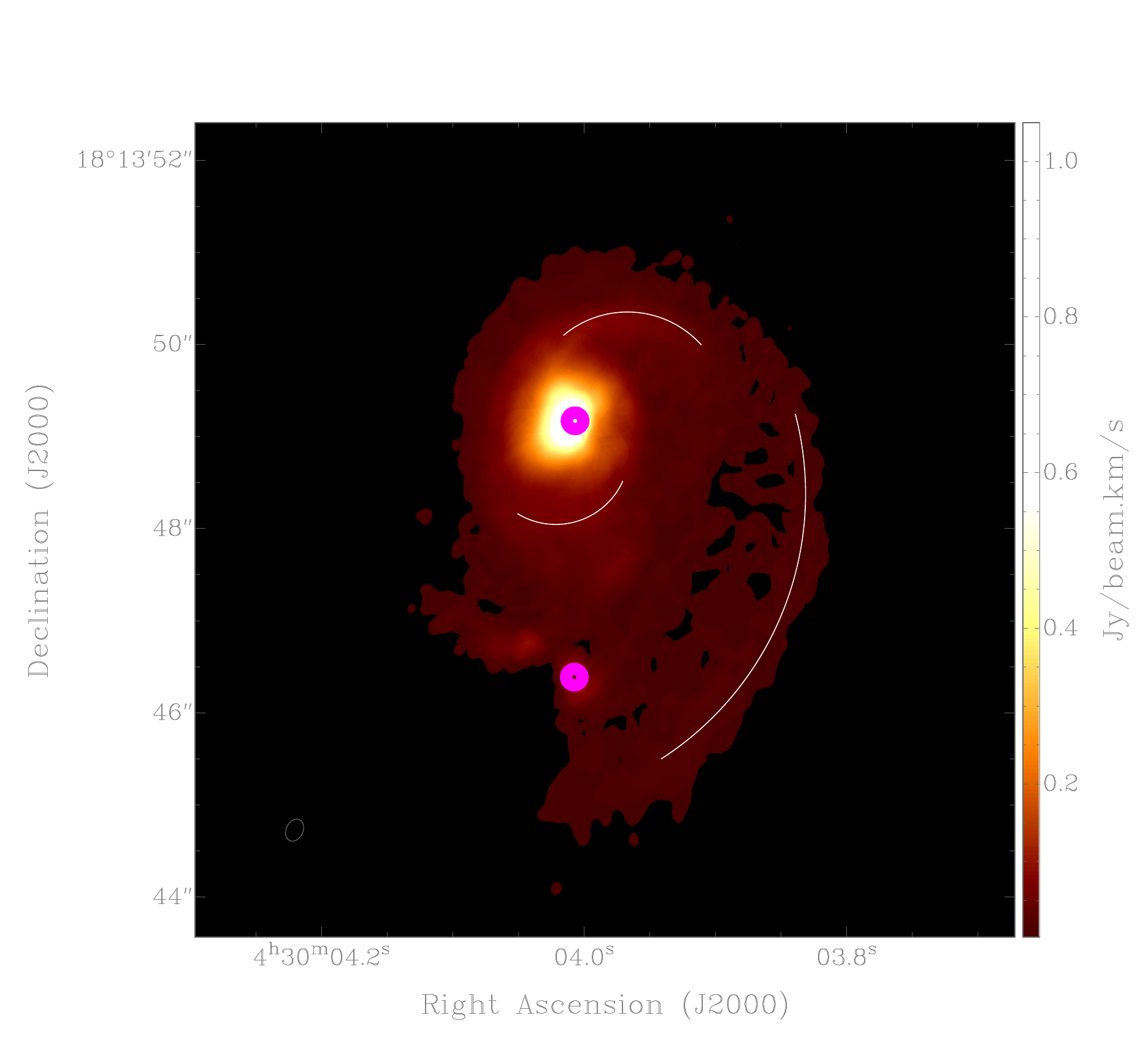}{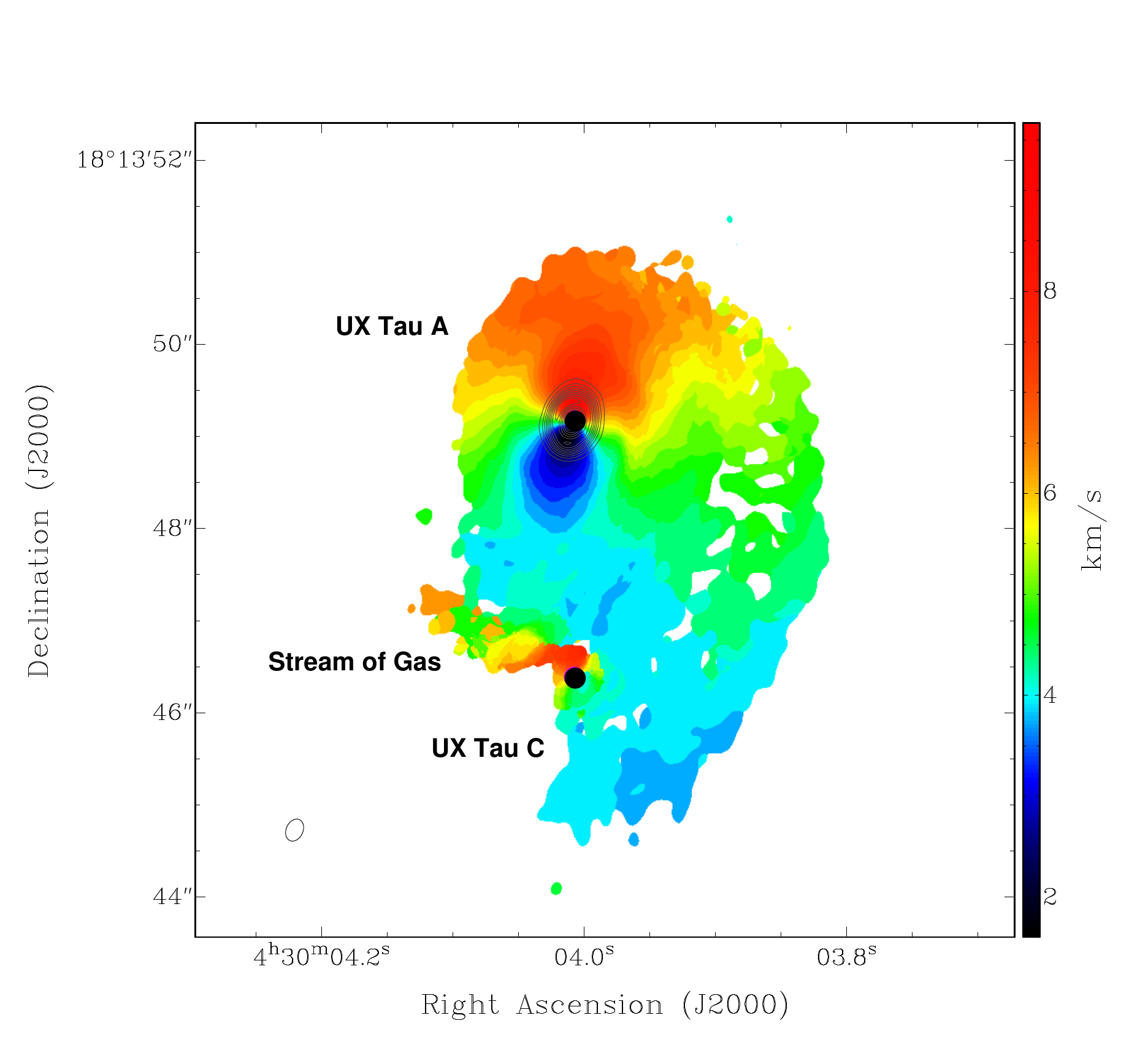}
\vspace{-1cm}
\caption{  ALMA CO(3$-$2) moment zero (left)  and one (right) images from UX Tauri A and C star system. In the right image (the moment one), 
we have additionally overlaid in grey contours the continuum 0.85 mm emission from UX Tauri A.  We do not detect continuum emission associated with UX Tauri C. 
The contours range from 10\% to 90\% of the peak emission,
in steps of 10\%. The peak of the millimeter continuum emission is 25 mJy beam$^{-1}$. 
The half-power contour of the synthesized beam of the line image is shown in the bottom-left corner.
We integrated in radial velocities from $-$3.8 to $+$13.8 km s$^{-1}$  for the CO gas disks in UX Tauri.
The LSR systemic velocity of the UX Tauri system is about $+$5.1 km s$^{-1}$. The black and magenta circles trace the position 
of the optical stars reported by the GAIA DR2 data obtained from the SIMBAD database, we have additionally
corrected by the proper motions for both sources. The scale bars in the right represent the intensity (left image) 
and the radial velocities (right image).  The white lines on the right image mark the most evident spiral arms.
The position of the component UX Tauri B is outside of this Figure.
\label{fig:f1}}
\end{figure*}
 
 \section{Observations} \label{sec:obs}
 
 \subsection{Band 7} \label{sec:band7}
 
 The archive observations of UX Tauri were carried out with ALMA in 2016 August 10th and 13th as part of the Cycle 4 program 2015.1.00888.S (PI: Akiyama, Eiji).
 Part of these data (the 0.87 mm continuum emission) have been already published in \citet{fran2020}. 
 The observations used 39 (in August 10th) and 37 (August 13th) antennas with a diameter of 12 m, yielding baselines with projected lengths 
 from 15 $-$ 1500 m (18.7 $-$ 1875 k$\lambda$).  The primary beam at this frequency has a full width at half-maximum (FWHM) of about 20$''$,  
 so that UX Tauri was covered with a single pointing at the sky position  $\alpha(J2000) = 04^h~ 30^m~ 3\rlap.^s99$;
$\delta(J2000) = $+$18^\circ~ 13'~ 49.41''$. The integration time on source was about 27 min in both sessions.

The weather conditions were very good and stable for these observations, with an average 
precipitable water vapor between 0.1 and 0.6 mm and an average system temperature 
around 120 K for August 10th and 150 K for August 13th. The ALMA calibration included simultaneous 
observations of the 183 GHz water line with water vapor radiometers used to reduce atmospheric phase fluctuations.
The quasars J0510$+$1800 and J0431$+$2037 were used for the bandpass, atmosphere, pointing, water vapor radiometer, and flux calibrations. 
J0431$+$1731 was used for correcting the gain fluctuations. 

The continuum images were obtained by averaging line-free spectral channels of four
spectral windows centered at:  344.003 GHz(spw0), 355.545 GHz(spw1), 345.809 GHz(spw2), and 356.747 GHz(spw3). 
The total bandwidth for the continuum is about 4.9 GHz.  
These four spectral windows were centered to observe different molecular lines 
as the HCO$^{+}$(4$-$3) ($\nu_\mathrm{rest}$= 356.734223 GHz), and $^{12}$CO(3$-$2) ($\nu_\mathrm{rest}$=345.795989 GHz).
In this study, we concentrated only in the $^{12}$CO(3$-$2) spectral line and the continuum.  
Each window has a native channel spacing of 488.3 kHz or $\sim$0.41 km s$^{-1}$.

The data were calibrated, imaged, and analyzed using the Common Astronomy Software Applications (CASA)
Version 4.7. Imaging of the calibrated visibilities was done using the CLEAN and TCLEAN tasks. 
We concatenated the data from both dates with the CONCAT task.
We set the {\tt\string robust} parameter of CLEAN in CASA to natural for the CO and 0.5 for the continuum image.
 The resulting image rms noise for the continuum was 40 $\mu$Jy beam$^{-1}$ at an angular resolution of 0.18$''$ $\times$ 0.16$''$
 with a PA of $-$30$^\circ$. The ALMA theoretical rms noise for this configuration, integration time, and frequency is about 35 $\mu$Jy beam$^{-1}$,
 which is also very close to the value we obtained in the continuum images.  For the line image rms noise we obtained a value of 3.0 mJy beam$^{-1}$ km s$^{-1}$ at an angular 
 resolution of 0.23$''$ $\times$ 0.18$''$ with a PA of $-$24.6$^\circ$. 
The ALMA theoretical rms noise for this configuration, integration time, bandwidth (channel spacing), and frequency is about 4.3 mJy beam$^{-1}$,
 which is also very close to the value we obtain in the line images. 
 Phase self-calibration was done in the continuum images and some improvement was obtained. We then applied the solutions
 to the line maps, obtaining a significant improvement in the channel maps. 
 The thermal emission from the CO can be seen in Figures \ref{fig:f1}, \ref{fig:f3}, and \ref{fig:f4}.

 \begin{figure*}[ht!]
\epsscale{1.165}
\plottwo{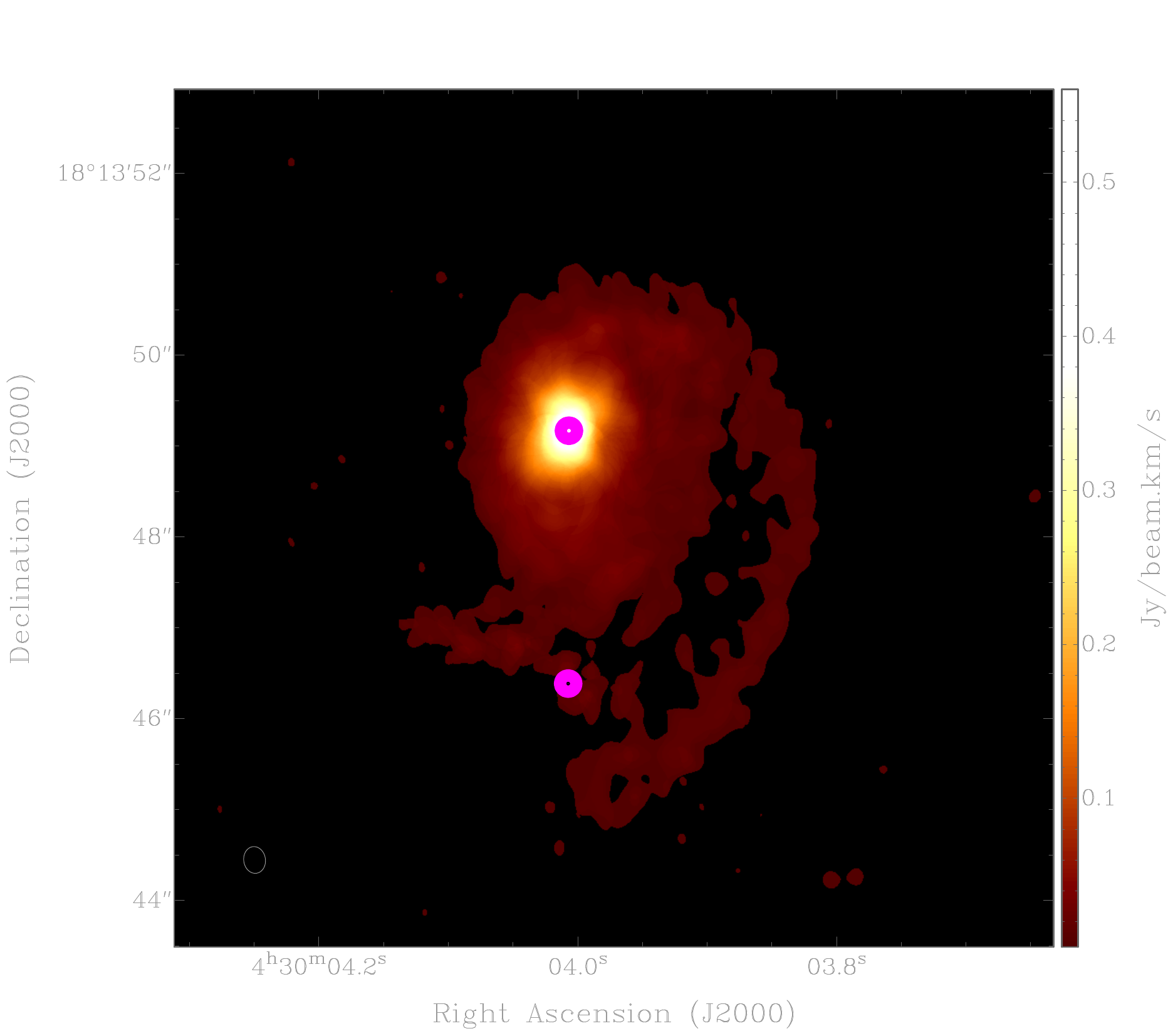}{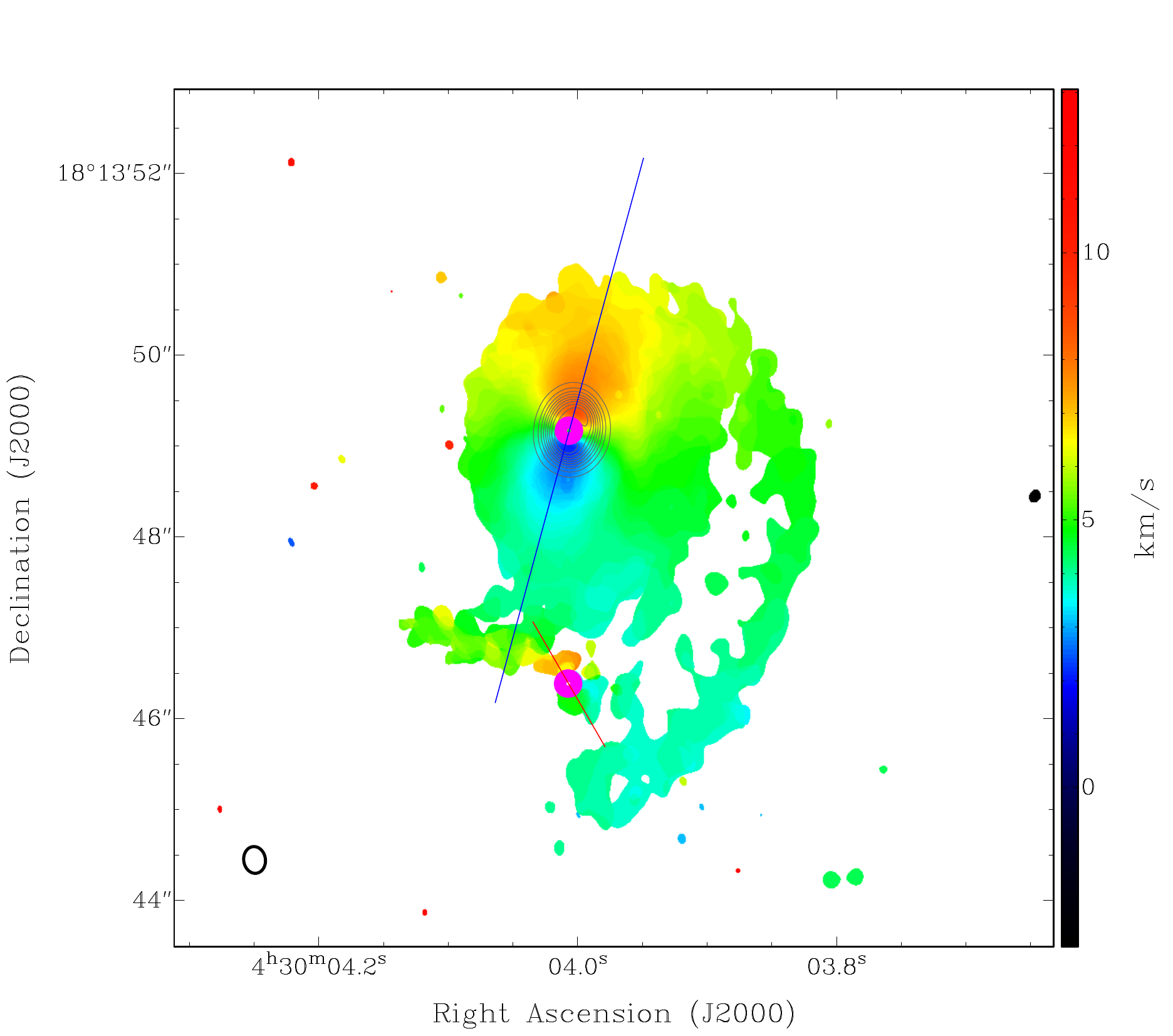}
\caption{ ALMA CO(2$-$1) moment zero (left)  and one (right) images from UX Tauri A and C star system. In the right image (the moment one), 
we have additionally overlaid in grey contours the continuum 1.3 mm emission from UX Tauri A. We do not detect continuum emission associated with UX Tauri C. 
The contours range from 10\% to 90\% of the peak emission,
in steps of 10\%. The peak of the millimeter continuum emission is 15 mJy beam$^{-1}$. 
The half-power contour of the synthesized beam of the line image is shown in the bottom-left corner.
We integrated in radial velocities from $-$3.8 to $+$13.8 km s$^{-1}$  for the CO gas disks in UX Tauri.
The LSR systemic velocity of the UX Tauri system is about $+$5.1 km s$^{-1}$. The black and magenta circles trace the position 
of the optical stars reported by the GAIA DR2 data obtained from the SIMBAD database, we have additionally
corrected by the proper motions for both sources, similar to Figure \ref{fig:f1}. The scale bars in the right represent the intensity (left image) 
and the radial velocities (right image).  The blue (UX Tauri A) and red (UX Tauri C) lines trace the position where the position-velocity 
diagrams shown in Figures \ref{fig:f4} and \ref{fig:f5} were computed.  The position of the component UX Tauri B is outside of this Figure.
\label{fig:f2}}
\end{figure*}

 \subsection{Band 6} \label{sec:band6}
       
The archive observations of UX Tauri were carried out with ALMA in 2015 August 12th as part of the Cycle 2 program 2013.1.00498.S (PI: P\'erez, Laura).
 Part of these data (the 1.3 mm continuum emission) have been already published in \citet{pini2018,fran2020}.  
 The observations used 44 antennas with a diameter of 12 m, yielding baselines with projected lengths 
 from 15 $-$ 1574 m (11.5 $-$ 1210 k$\lambda$).  The primary beam at this frequency has a full width at half-maximum (FWHM) of about 30$''$,  
 so that UX Tauri was covered with a single pointing at the sky position  $\alpha(J2000) = 04^h~ 30^m~ 4\rlap.^s00$;
$\delta(J2000) = $+$18^\circ~ 13'~ 49.20''$. The integration time on source was about 14.7 min.  

 The quasars J0423$-$0120, J0510$+$180, and J0431$+$2037 were used for the bandpass, atmosphere, pointing, water vapor radiometer, and flux calibration. 
 J0431$+$2037 was used for correcting the gain fluctuations.    

The continuum images were obtained by averaging line-free spectral channels of eight 
spectral windows. The total bandwidth for the continuum is about 6.3 GHz.  
These eight spectral windows were centered to observe different molecular lines 
as for example the $^{13}$CO(2$-$1) ($\nu_\mathrm{rest}$=  220.3986 GHz), and $^{12}$CO(2$-$1) ($\nu_\mathrm{rest}$= 230.5380 GHz).
Each window has a native channel spacing of 488.3 kHz or $\sim$0.61 km s$^{-1}$.
In this study, we concentrated only on the $^{12}$CO(2$-$1) spectral line and the continuum.     
        
The data were calibrated, imaged, and analyzed using the CASA
Version 4.4. Imaging of the calibrated visibilities was done using the CLEAN and TCLEAN tasks. 
 We set the {\tt\string robust} parameter of CLEAN in CASA to {\tt\string Briggs}$=$2.0 for the CO and the continuum images.
 The resulting image rms noise for the continuum was 46 $\mu$Jy beam$^{-1}$ at an angular resolution of 0.30$''$ $\times$ 0.24$''$
 with a PA of $+$9$^\circ$. The ALMA theoretical rms noise for this configuration, integration time, bandwidth, and frequency is about 34 $\mu$Jy beam$^{-1}$,
 which is very close to the value we obtain in the continuum images.  For the line image rms noise we obtained a value of 3.2 mJy beam$^{-1}$ km s$^{-1}$ at an angular 
 resolution of 0.29$''$ $\times$ 0.24$''$ with a PA of $+$8.2$^\circ$. 
The ALMA theoretical rms noise for this configuration, integration time, bandwidth (channel spacing), and frequency is about 3.0 mJy beam$^{-1}$,
 which is very close to the value we obtain in the line images. 
 Phase self-calibration was done in the continuum images, obtaining again a significant improvement in the channel maps.         
 The thermal emission from the CO can be seen in Figure \ref{fig:f2} and \ref{fig:f5}.

 \section{Results and Discussion} \label{sec:res}
 
 \subsection{UX Tauri System (A/C)}

In Figures \ref{fig:f1} and \ref{fig:f2} we present the main results of this ALMA study. In Figure \ref{fig:f1}, the moment zero (integrated intensity) and one 
(intensity-weighted velocity) maps of the CO(3$-$2) thermal emission are presented. In addition to these maps, we have overlaid in contours 
the 0.85 mm continuum emission together with the moment one map, and the positions of the optical stars related with UX Tauri A and C obtained with the GAIA 
DR2 data and corrected by the proper motions obtained from this catalogue \citep{gai2018}, see Figure \ref{fig:f1}.  
The continuum images are very similar  to those presented in \citet{pini2018,fran2020}. 

The high fidelity and sensitive CO images of the UX Tauri A/C system presented here reveal for the first time the gas structure of the disks. 
Recent ALMA images of UX Tauri A/C could not reveal this structure and only imaged the molecular emission very close to the UX Tauri A disk, see 
Figure 4 from \citet{ake2019}.  The line rms-noise obtained for the CO(2$-$1) in 
\citet{ake2019} was however very high (39 mJy beam$^{-1}$ km s$^{-1}$) compared with our observations (3.2 mJy beam$^{-1}$ km s$^{-1}$). 
On the other hand, our maps reveal for the first time marked molecular gas spiral arms connecting the UX Tauri A/C system.  
These arms seem to arise from the north side of UX Tauri A and go directly to the south, where its companion C is localized. 

Futhermore, there are two thin and bright (with an angular size of only about 1$''$) 
spiral arms close to UX Tauri A, one begins in the north part of the disk and extends to the west, and the other one starts in the south and extends to the east (see Figure 1).  
Very close to the innermost part of the UX Tauri A disk, where the continuum emission is located, the CO emission is probably optically thick, and not much structure is observed.
This physical opacity effect has already been observed in other disks as for example in IRAS16293 \citep{zap2013} at higher frequencies, and HL Tau \citep{alm2015}.     

In Figure \ref{fig:f1}, as mentioned before, we include a map of the CO(3$-$2) moment one emission from the UX Tauri system. In order to compute this map,
we integrated in radial velocities from $-$3.8 to $+$13.8 km s$^{-1}$  for the CO gas disks in UX Tauri. The LSR systemic velocity of the UX Tauri system 
is about $+$5.1 km s$^{-1}$ \citep{ake2019}. From this image, we can see clearly the two rotating disks associated to UX Tauri A and C. In both disks the redshifted velocities 
are found to the north and the blue-shifted ones are toward the south. There is one extra-tail that seems to connect the redshifted part of the rotating disk associated 
with UX Tauri C. The redshifted velocities begin from the UX Tauri C disk and go to the east, with a pattern clearly different from the Keplerian behavior of the gas in the UX Tauri A disk.   
This is the first time that this disk and the tail is reported. The kinematics of the gas revealed in Figure \ref{fig:f1} shows that the radial velocities of the spiral tails in this system follow the same structure
as the disk in UX Tauri A, which suggests that these tails are likely disrupted molecular material from UX Tauri A surroundings.     

From Figure \ref{fig:f1}, we can obtain the positions and structure of the disks. For the UX Tauri A CO molecular disk, we obtained from our model a deconvolved size 
 1800 milliarcsec with a Position Angle of 164$^\circ$ $\pm$ 8$^\circ$.  For the disk in UX Tauri C,
it is more difficult to obtain a deconvolved size and a PA because there is a large amount of intervening gas from the eastern tail and the UX Tauri A spiral arms. 
However, from Figure \ref{fig:f1} we estimated roughly a size of $<$400 milliarcsec and a PA of 30$^\circ$. We should be cautious with these numbers obtained 
for the component C because they are upper limits.  As the position angles are different for both disks, this suggests that maybe the disks are not co-planar 
(or are not rotating in the same plane) and the eastern molecular tails are also in a different plane.

 \begin{table*}[]
\begin{center}
Physical Parameters from the Best Keplerian Thin Disk Fit for UX Tauri A\\ 
\bigskip
\begin{tabular}{lc}
\hline
\hline
Physical Parameter~~~~~   &  Estimated Value from the Model  \\        
\hline
Line-width ($\Delta v$)~~~~~            &  1.6$\pm$0.2 km s$^{-1}$  \\
Beam-width ($\Delta s$)~~~~~          &  0.21 arcsec (fixed)\\
Central velocity of the Disk (v$_0$)   &  $+$5.1$\pm$0.1 km s$^{-1}$\\ 
Inclination Angle ($i$)                         &  32 $^\circ$ $\pm$ 8$^\circ$\\ 
Inner Disk Radius ($r_1$)                       &  0.01$''$(fixed)\\
Outer Disk Radius ($r_2$)                      &  1.8$''$$\pm$0.3$''$\\
Infall Velocity (v$_i$ at $r_0$)              &  $-$0.1$\pm$0.2 km s$^{-1}$\\
Rotational Velocity (v$_r$ at $r_0$)     & $+$4.5$\pm$1.0 km s$^{-1}$\\
Reference Radius ($r_0$)                    & 0.5$''$ (70 au)\\
Enclosed Mass (star+gas)                    & 1.4 $\pm$ 0.6 M$_\odot$\\
\hline
\hline
\end{tabular}
\label{tab:t1}
\end{center}
\end{table*}   

\begin{figure*}
\centering
\vspace{-2cm}
\includegraphics[scale=0.6, angle=0]{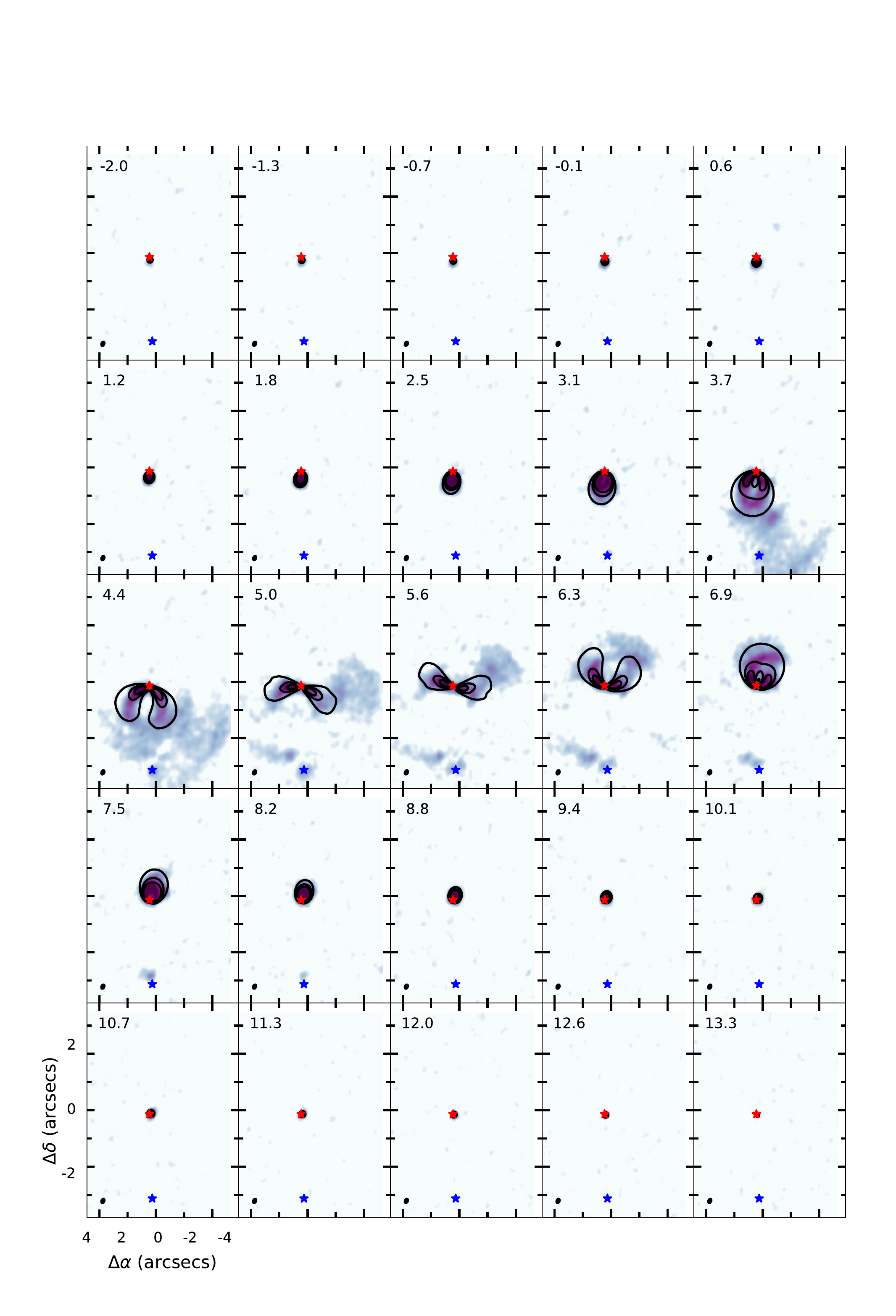}
\caption{ ALMA CO(3$-$2) channel velocity maps overlaid with our best synthetic model of a Keplerian thin disk in black contours of UX Tauri A. 
The contours are in percent of the peak emission. 
The red and blue stars mark the position of the optical stars associated with UX Tauri A
and UX Tauri C, respectively. The radial velocity corresponding to each channel bin is shown in the upper-left corner and is in km s$^{-1}$. Every channel map
is centered in UX Tauri A. Here, we smoothed our channel spacing to 0.7 km s$^{-1}$ to obtain a better sensitivity.   
 The position (0,0) corresponds to the position of UX Tauri A. 
\label{fig:f3}}
\end{figure*}    

For the continuum emission, our image reveals the pre-transitional disk reported for UX Tauri A and already mapped at these wavelengths 
by the SMA and ALMA \citep{and2011, ake2019}, as it is mentioned earlier. 
 Our continuum 0.85 mm map shows a similar structure as in these observations, as for example the inner cavity.
 We do not detect UX Tauri C component at a 5$\sigma$-level (0.2 mJy beam$^{-1}$).  
 
Figure \ref{fig:f2} is similar to Figure \ref{fig:f1}, but with the line CO(2$-$1) detected at band 6.  Here, we have additionally included the 1.3 mm continuum 
emission from the UX Tauri A/C system. The thermal CO(2$-$1) and 1.3 mm continuum emission show a very similar structure to that revealed in the CO(3$-$2) 
and 0.85 mm images as shown in Figure \ref{fig:f1}. 
This Figure confirms the marked molecular gas spiral arms connecting the UX Tauri A/C system, and the two rotating disks associated with UX Tauri A and C. These rotating disks
have similar characteristics as revealed in the CO(3$-$2) maps. From both Figures, the CO molecular emission from the disk of UX Tauri A has an extent of about 3$''$ $\times$ 2$''$ 
or 420 au $\times$ 280 au, quite large compared to the (sub)millimeter continuum disk (a factor of $>$ 5). 
 Even if compare this difference in size between the continuum and the molecular emission in other transitional disks \citep[{\it e.g.} 
SR24S ($\sim$ 1); 2MASS J16042165-213028 ($\sim$ 2); DM Tauri ($\sim$ 1); 
T Chamaeleontis ($\sim$ 2)][]{fer2017, may2018, ter2019, kud2018,hue2015} this difference is still quite large for UX Tauri A.
However, \citet{fan2019} reported in CX Tauri a large difference of the millimeter dust and gas extents ($>$ 5) and argue that the this could be explained due to   
a radial drift, and closely matches the predictions of theoretical models. 
This difference in dust and gas extents is notorious for UX Tauri A, however, it is not clear for UX Tau C because we did not detect the continuum emission. 
 
Figure \ref{fig:f3} presents the channel spectral maps of the CO(3$-$2) emission from the UX Tauri A/C system.  The channel maps from the CO(2$-$1) emission 
are similar and are not presented here.  Together with this CO(3$-$2) channel map we have computed a Keplerian thin disk model fitted to the line data from UX Tauri A and overlaid it 
in this channel map. As the emission from UX Tauri C is very contaminated by the disk interaction between both components, a Keplerian thin disk model could not
be obtained.  \citet{pin2018} demonstrated that neglecting the vertical structure of the disk (assuming a thin disk) has a small impact on the derived stellar mass, especially in CO lines.
Thus, the dynamical mass and other parameters obtained with this simple model constrain well the disk structure.

The best physical parameters of the Keplerian disk surrounding UX Tauri A are presented in Table \ref{tab:t1}. 
This Keplerian thin-disk model is described in detail  in the Appendix B of \citet{zap2019} and we will not discuss this again here. From Figure \ref{fig:f3}, we can note that there
is good correlation between the channel map emission from the disk of UX Tauri A and the best Keplerian thin disk model fit.   
Residual maps from the model and observations can be obtained, however, there is a large amount of intervening emission that gets bright in the residual,
thus that we are not presenting these maps here.  From this disk model, we obtained three important parameters with a good precision, the inclination angle {\it i} = 32$^\circ$ $\pm$ 8$^\circ$, 
the size 1.8$''$$\pm$0.3$''$ (250 au) and the enclosed mass (disk$+$star) 1.4 $\pm$ 0.6 M$_\odot$. 
The compact gas disk from UX Tauri C and the red-shifted molecular gas bridge can also be seen in the southern part of the channel maps. 

\begin{figure}
\includegraphics[scale=0.53]{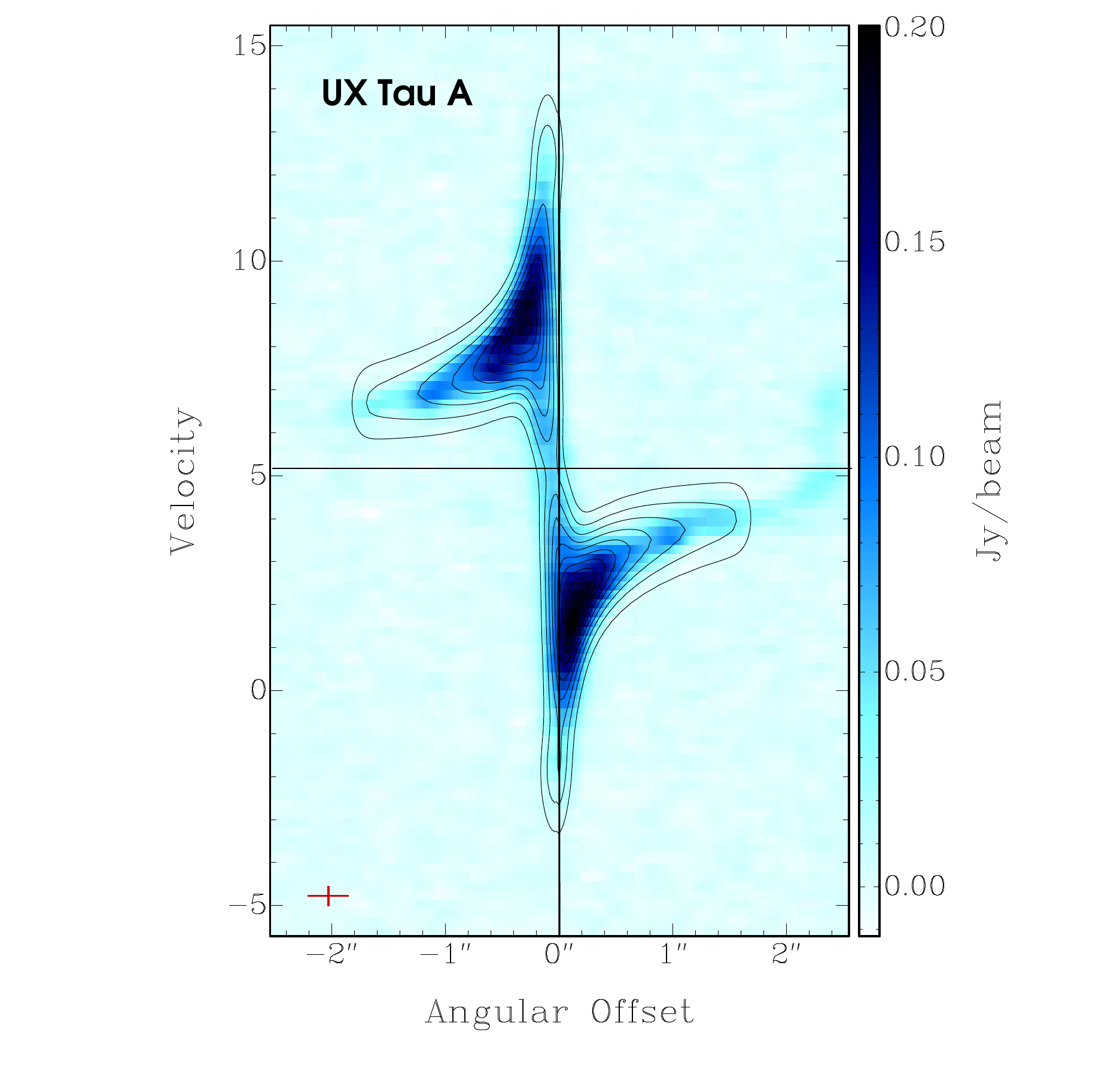}
\caption{Position-velocity diagram computed along the major axis of the circumstellar disk of UX Tauri A (PA=164 deg). The blue scale image
is obtained from the ALMA CO(3$-$2) data and the contour image is from our synthetic model.  
The contours range from 10\% to 90\% of the peak emission,
in steps of 10\%. The peak of the millimeter continuum emission is 0.2 Jy beam$^{-1}$. 
The spectral and angular resolutions are shown in the bottom-left corner. The black lines mark the systemic cloud velocity ($+$5.1 km s$^{-1}$)
and the disk axis center. The scale bar in the right represents the intensity.
\label{fig:f4}}
\end{figure}

\begin{figure}
\centering
\includegraphics[scale=0.77, angle=0]{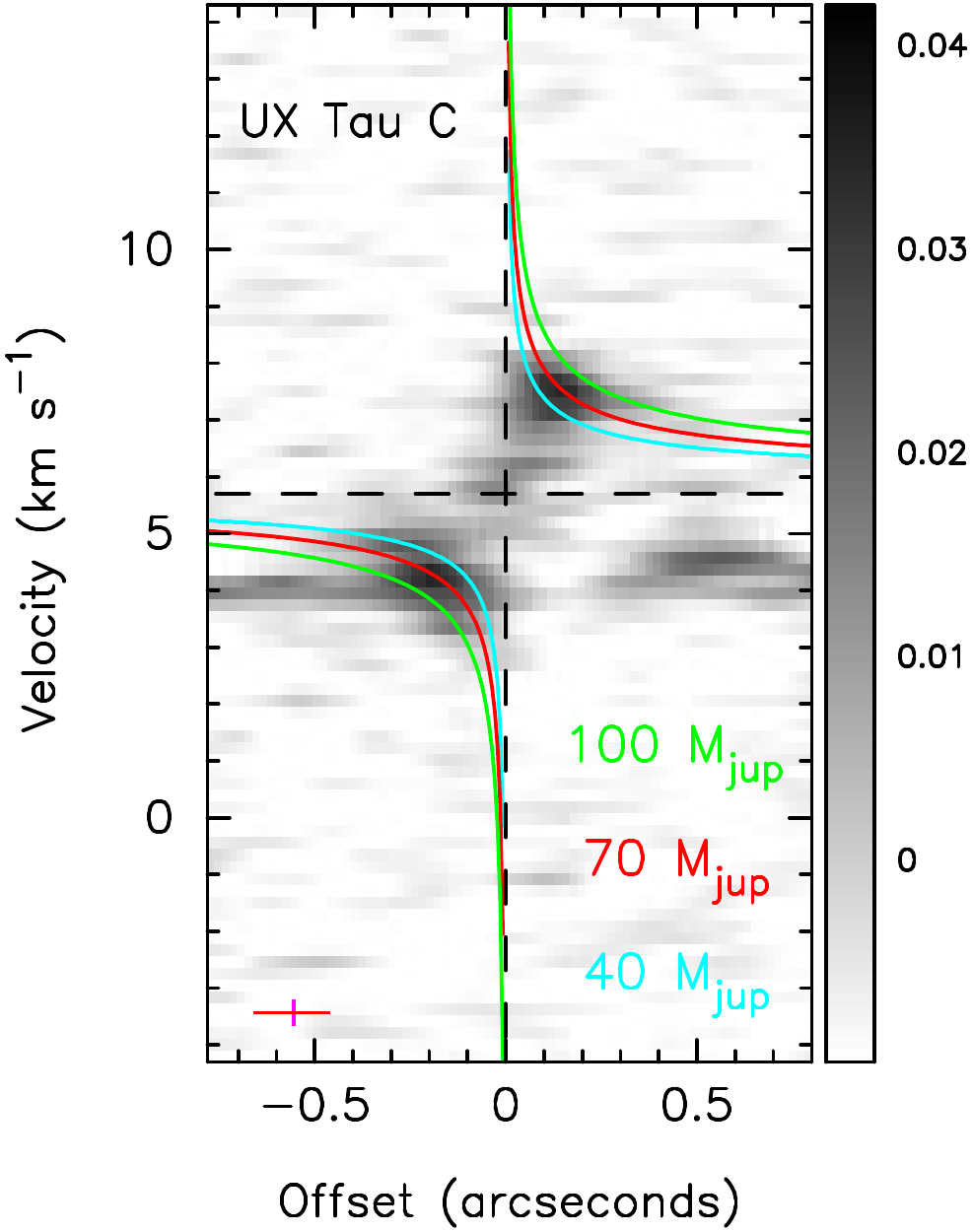}
\caption{Position-velocity diagram computed along the major axis of the circumstellar disk of UX Tauri C (PA=30 deg). The grey scale image
is obtained from the ALMA CO(3$-$2) data.  
The spectral and angular resolutions are shown in the bottom-left corner. The dashed lines mark the systemic cloud velocity ($+$5.8 km s$^{-1}$)
and the disk axis center. The scale bar in the right represent the intensity and is in Jy beam$^{-1}$.
The cyan, red, and green lines mark the Keplerian rotation curves for a disk with an enclosed masses (disk$+$star) of 100 (white), 
 70 (red), and 40 (green) Jupiter masses, respectively. 
\label{fig:f5}}
\end{figure}

Figures \ref{fig:f4} and \ref{fig:f5} reveal the kinematics of the molecular gas within both disks, UX Tauri A/C. In these Figures, we present the Position-Velocity (PV) diagrams
obtained along the major axis of both disks. In the PV diagram of the disk of UX Tauri A (Figure \ref{fig:f4}), we have additionally overlaid the PV diagram from the thin disk model obtained
in this paper. Both PV diagrams show a good correspondence, revealing clearly and with a good fidelity the Keplerian curves.  
For the case of UX Tauri C, we could not obtain a model, so we only plotted the rotation Keplerian curves
of enclosed masses (disk$+$star) of 100, 70, and 40 Jupiter masses.  Therefore, we consider that UX Tauri C could have an enclosed mass of 70 $\pm$ 30 / $\sin i$ (where $i$ is the inclination) 
Jupiter masses, on the limit
between a brown dwarf and a low-mass star.  
Here, we did not correct the mass for the inclination of the disk because the gas emission from the disk is very contaminated. ALMA observations with a better angular resolution 
might help in better tracing the disk and thus its inclination angle.  The physical size of this gas disk, is $<$ 56 au, as mentioned earlier. 
As this gas disk is first reported here, these parameters cannot be compared, however, we note that the mass of the star obtained from the optical 
(160 Jupiter masses) \citep{kra2009}.  More recent estimations however reported a mass for UX Tauri C 
of 87 Jupiter masses \citep{cse2017}, which is very similar to that given here.
These masses however were obtained from the stellar models, contrary to our values that are obtained from the ALMA CO observations.


\subsection{Dust and Gas Mass}

From a Gaussian fitting to the 0.85 mm continuum emission from UX Tauri A, we obtained a deconvolved size of 570$\pm$30 $\times$ 420$\pm$30 milliarcsec$^2$
with a PA of 167$^\circ$ $\pm$ 9$^\circ$ and an integrated flux of 197 $\pm$ 12 mJy. 
The position angle is in good agreement with the one obtained for the CO(3-2) (see above).
The corresponding physical sizes of these deconvolved values are about 70~au, 
which suggest we are only seeing the circumstellar disk and at a level of 5$\sigma$ (0.175 mJy Beam$^{-1}$) no dust emission arises from the spirals measured in CO emission.
Assuming that the dust emission is optically thin and isothermal, the dust mass (M$_\mathrm{d}$) is directly proportional to the flux density (S$_\nu$)
integrated over the source, as:

\begin{equation}
M_d=\frac{D^2 S_\nu}{\kappa_\nu B_\nu(T_d)},
\label{eq1}
\end{equation}

\noindent
where $D$ is the distance to UX Tauri A  \citep[139 pc,][]{bai2018}, $\kappa_\nu$ the dust mass opacity, and B$_\nu(T_d)$ the Planck function for the dust
temperature T$_d$.  In reality, the thermal dust emission is probably not optically thin, hence the estimated mass is considered to be a lower limit. 
Assuming a dust mass opacity ($\kappa_\nu$) of 0.025 cm$^2$ g$^{-1}$ (of dust$+$gas, and taking a gas-to-dust ratio of 100) appropriate for these wavelengths (1.3 mm) \citep{Oss1994}, 
as well as a characteristic dust temperature of 50 K, we
estimated a lower limit for the mass (dust$+$gas) of the disk associated with UX Tauri A of 0.005 M$_\odot$. The mass uncertainty could be very large (a factor of up to 3 or 4) 
given the uncertainty in the opacity and in the dust temperature. This mass estimation is in good agreement with the value estimated by \citet{and2011}.
For the component UX Tauri C, we can estimate an upper limit for its total dust mass. Taking a dust temperature of 20 K, because this is a young star with a low
mass, a flux density of 0.2 mJy,  and a dust mass opacity similar to that of UX Tauri A, we found a dust mass for UX Tauri C of $\leq$ 0.05 M$_\Earth$. 
This dust mass could be associated with the disk of UX Tauri C. A similar upper limit can be considered for the component UX Tauri B.

Molecular disks have been detected toward a handful of low mass or substellar objects. 
Some examples are IC 348-SMM2E \citep{pal2014}, TWA 34 \citep{rod2015},  Rho Oph 102 \citep{ric2012} and 2MASS J04442713+2512164 \citep{ric2014}. 
However, none of these disks seem to be associated with a multiple system as UX Tauri.

\subsection{UX Tau B}

We did not detect continuum or line emission from 
the companion UX Tauri B in both bands, localized to the west, above of a 5$\sigma$ threshold 
(200 $\mu$Jy beam$^{-1}$ for the continuum at 0.85 mm, and 15  mJy beam$^{-1}$ km s$^{-1}$ for the CO(3-2) line). 
The disk of UX Tau B falls outside of the images and it is thus not indicated.


\subsection{UX Tauri C: A Possible Low-Mass Object Flyby}
    
 The ALMA CO molecular observations toward the UX Tauri system clearly reveal the tidal interaction between the components of the UX Tauri A/C system, exhibiting pronounced spiral arms.
 This type of physical interaction in young triple and higher order multiple stellar systems have been discussed in \citet{rei2014}. 
  Moreover, this scenario was already debated by \citet{cse2017} for this system, where they analyzed multi-epoch high-spatial resolution observations in 
 optical and NIR bands in UX Tauri and concluded that despite UX Tauri A and B are being probably gravitationally bound, the case of  UX Tauri A and C system is not entirely clear.
 However, since the semi-major axis of the orbit between UX Tauri A and C is significantly affected by the estimation of the inclination, they could not confirm the classification of UX Tau A/C as a  
 unstable triple system.
 
 We propose that UX Tauri C is maybe experiencing a flyby due to a close-by dynamical interaction with the component UX Tau A. It is likely that this flyby is
 not a one-time flyby by a completely unaffiliated random young brown dwarf in a similar evolutionary state that happens to be free-floating 
 in the sparse Taurus association. Therefore, UX Tau A/C could be experiencing a closest approach of a wide, evolving and eccentric orbit for 
 the first time in thousands of years.
 
 Here, we mention some characteristics from our new ALMA observations of the system supporting 
 this flyby scenario: 

\begin{itemize}
	 \item The two CO disks are probably not coplanar nor aligned. 
	 \item  The CO extended gas shows spiral arms and a bridge linking the A and C disks. 
	   These features are predicted by simulations of prograde close encounters or flybys \citep[{\it e.g.}][]{vor2017,cue2019,cue2020}. 
	 \item The spiral arms in UX Tauri A disk have pitch angles between $20\degr$ and $30\degr$. These large values do not agree with 
	   the spiral patterns produced by gravitational fragmentation of disks, which are typically more roundish, with pitch angles ranging $10\degr-20\degr$ \citep{cue2019,cue2020}. 
	   The pitch angles of the disk interactions seem to be small like in the case of AS 205 \citep[14$^\circ$;][]{kur2018A} or for HT Lup A \citep[4$^\circ$;][]{kur2018A}.
	    
	\item The kinematics of the CO tail extending due east of the disk of UX Tauri C are compatible with a structure non-coplanar with the UX Tauri A disk. 
	  It is at a more redshifted radial velocity than the gas from the south part of the UX Tauri A disk. It shows a velocity gradient from west (blue) to east (red), 
	  bridging some of the southwest part of the UX Tauri A disk with the redshifted part of the UX Tauri C disk. Hence, this CO bridge could be interpreted as a loop 
	  of gas in the foreground of the UX Tauri A disk, rising from this disk and infalling into the UX Tauri C disk. This kind of loop structures are produced in simulated 
	  encounters between a disk and a prograde perturber with a trajectory inclined with respect to the plane of the main disk \citep[{\it e.g.} Figures 4 and 5 for a flyby 
	  with inclination $45\degr$ in][]{cue2019,cue2020}.
	\item UX Tauri C has a gas disk. Simulations of prograde flybys also predict that the perturber strips gas from the main disk \citep{vor2017,cue2019,cue2020}.
	\item The UX Tauri A disk presents a large central cavity. In addition to the truncation of the disks a flyby can also produce wide inner cavities through tidal 
	  interactions at the core of the perturbed disks as shown again in the recent flyby simulations by \citet{cue2019,cue2020}. 
\end{itemize}

There could exist other possible scenarios to generate spiral arms in a disk, but they are not as favourable as the flyby described before. Either disk fragmentation
 due to gravitational instabilities, perturbations by a planet or a stellar embedded companion, or a massive external body via gravity or asymmetric illumination 
 may induce the formation of spiral arms in a disk. However, a disk fragmentation is not favourable in disks with small masses like the one around UX Tauri A. 
 We can make a rough estimate of the Toomre Q parameter, which for Keplerian disks is approximately:
 
 \begin{equation}
 Q\simeq2(M_*/M_{disk})\cdot(H/R_{disk}), 
 \end{equation} 
 
  where $M_*$ and $M_{disk}$ are the mass of the star and the disk respectively, $H$ is its scale height and $R_{disk}$ its outer radius. Assuming a low average 
 disk temperature of 50\,K (hence a sound speed $c_s=0.6$ km s$^{-1}$), an angular velocity $\Omega$ = 4 $\times$ 10$^{-10}$ km s$^{-1}$ (derived using the values of v at $r_0$ from Table 2), 
 the outer disk radius of 252 \,au (Table 2) and the derived disk and stellar masses (0.007 M${_\odot}$ and 1.4 M${_\odot}$, respectively), we obtain a $Q=11$, well over unity, implying that 
 UX Tauri A is a stable disk against gravitational disturbances. An embedded companion in the disk typically displays a spiral pattern with smaller pitch angle and a 
 more symmetric pattern, with equal spirals in both sides of the disk.  An external perturber (farther out than UX Tau C) is also ruled out, since UX Tauri B is probably not massive nor luminous 
 enough to induce the spirals displayed by the disk. It is important to mention that the spiral arms connect directly to UX Tau C, discarding UX Tau B as a perturber.  

One can estimate if UX Tauri C is bounded to UX Tauri A gravitational field. From Figure 5 in \citet{cse2017}, we estimated a tangential velocity of 0.1$''$ per 27.5 yrs    
or 2.4 $\pm$ 0.5 km s$^{-1}$. This motion comes mainly from the DEC as the motion in RA is null. From the PV diagrams shown in Figures \ref{fig:f4} and \ref{fig:f5} we roughly 
estimated a radial velocity for UX Tauri C of about 0.7 $\pm$ 0.3 km s$^{-1}$. Thus, the total velocity is $\sim$ 2.5 $\pm$ 0.6 km s$^{-1}$.  
For a body orbiting a star with the mass of UX Tauri A (1.3 M$_\odot$) at a
distance of 360 au and in a circular orbit, it is needed an orbital velocity of less than 1.8 km s$^{-1}$ to be bounded. As the UX Tauri C velocity is a bit larger, this may indicate that this object is 
un-bounded to the gravitational field of UX Tauri  A, but if we take into account the uncertainties on the total velocity, it still is not very clear. 
Furthermore, if we take a parabolic orbit (which is expected from a periastron passage in a dynamically evolving eccentric and wide orbit), 
the orbital velocity would increase by a factor of $\sqrt 2$, so this might indicate that it is still bounded. 

All of the observational evidence presented here lead us to propose that UX Tauri C is passing in a prograde flyby close to the disk of UX Tauri A. The former UX Tauri C, 
coming from the background, has probably captured some gas from the UX Tauri A disk and it is forming a new disk which is already orbiting it. 
The tidal interaction has caused the formation of spiral arms, one of which is rotating in the plane of the main disk and the other is inclined 
toward the foreground, bridging both young stars.   

The case presented here is therefore  one of a few cases (HD 100453 AB, AS 205, SU Aurigae, RW Aurigae) of binary disk interactions mapped in molecular 
gas \citep{van2019, kur2018A, aki2019, dai2015}. AS 205 and RW Aurigae are considered as tidal encounters with parabolic orbits or flyby's. 
However, the intruder does not have a substellar mass in any of them, so this is the first time that it has been reported.
These two examples revealed molecular gas bridges between the disks and the stream-like structures extending from the main component. 
Depending on whether or not the encounter is prograde or retrograde the material transferred from one circumstellar disk to the other is more pronounced \citep{dai2015}.    
As mentioned before we then think that the encounter in UX Tauri A/C could be prograde, as suggested for the AS 205 object. 

\section{Conclusions} \label{sec:con}

 In this study, we present sensitive and high angular resolution ($\sim$0.2-0.3$''$) (sub)millimeter (230 and 345 GHz) continuum 
and CO(2$-$1)/CO(3$-$2) line observations of the UX Tauri A/C disk system located in the Taurus molecular cloud using
ALMA.  The main results of our work can be summarized as follows.
 
 \begin{itemize}
 
 \item These observations reveal the gas and dusty disk surrounding the young star UX Tauri A with a great detail, 
and for the first time it is detected the molecular gas emission associated with the disk of UX Tauri C (with a size of $\sim$ 56 au). No (sub)millimeter continuum
emission is detected at 5$\sigma$-level (0.2 mJy at 0.85 mm) associated with UX Tauri C in both bands.
 
 \item  We report a strong tidal disk interaction between both disks UX Tauri A/C, separated (in a projected distance) by only 360 au.
The CO line observations revealed marked spiral arms in the disk of UX Tauri A  and an extended redshifted stream of gas
associated with the UX Tauri C disk. 
We discuss the possibility of UX Tauri C being a young brown dwarf dwarf flyby given all the observational constrains. 

 \item  The enclosed masses (disk$+$star) estimation from the disk Keplerian radial velocities are 1.4 $\pm$ 0.6 M$_\odot$ for UX Tauri A, and 70 $\pm$ 30 / $\sin i$ Jupiter masses for UX Tauri C.
 For the component UX Tauri C, we can estimate a dust mass in its disk of $\leq$ 0.05 M$_\Earth$. 

 \end{itemize}

\facilities{ALMA}
\software{CASA\citep{mac2007}, KARMA\citep{goo1996}}

\acknowledgments

This paper makes use of the following ALMA data: ADS/JAO.ALMA\#2015.1.00888.S, and ADS/JAO.ALMA\#2013.1.00498.S. ALMA is a partnership of ESO (representing its member states), 
NSF (USA) and NINS (Japan), together with NRC (Canada), MOST and ASIAA (Taiwan), and KASI (Republic of Korea), in cooperation 
with the Republic of Chile. The Joint ALMA Observatory is operated by ESO, AUI/NRAO and NAOJ.
This research has made use of the SIMBAD database, operated at CDS, Strasbourg, France. L.F.R., is grateful to CONACyT, M\'exico, and DGAPA, 
UNAM for the financial support.   L.A.Z. acknowledges financial support from CONACyT-280775 and UNAM-PAPIIT IN110618 grants, M\'exico.
A.P. acknowledges financial support from CONACyT and UNAM-PAPIIT IN113119 grant, M\'exico.
GA and MO acknowledge financial support from the State Agency for Research of the Spanish MCIU through the AYA2017-84390-C2-1-R grant (co-funded by FEDER) and through the 
``Center of Excellence Severo Ochoa'' award for the Instituto de Astrof\'\i sica de Andalucia (SEV$-$2017$-$0709). 
 We  are  very  thankful  for  the  thoughtful  suggestions of  the  anonymous  referee and the editor that  helped  to  improve our manuscript.

%


\newpage

\bibliographystyle{aasjournal}



\end{document}